\documentclass[prd,twocolumn,showpacs,preprintnumbers,amsmath,amssymb,superscriptaddress]{revtex4}
\usepackage{graphicx, epsfig, rotating, dcolumn, bm}

\usepackage{atlasphysics}
\usepackage{color}
\def\s1{\hat{s}_1}
\def\T1{\hat{t}_1}
\def\t2{\hat{t}_2}
\def\U1{\hat{u}_1}
\def\u2{\hat{u}_2}

\begin{document}
%%%%%%%%%%%%%%%%%%%%%%%%%%%%%%%%%%%%%%%%%%%%%%%%%%%%%%%%%%%%%%%%%%%%%%%%%%%%%%%

%\preprint{arXiv:yymm.nnnn}
%\preprint{TIFR-TH/12-07}
\preprint{WITS-CTP-153}

\title{Probing the Higgs Boson via VBF with Single Jet Tagging at the LHC}
%%%%%

\author{Amanda Kruse}
\affiliation{Department of Physics, University of Wisconsin, 
Madison, WI 53706, USA. }
\author{Alan S. Cornell}
\affiliation{National Institute for Theoretical Physics; School of Physics, University of the Witwatersrand, Wits 2050, South Africa. }
\author{Mukesh Kumar}
\affiliation{National Institute for Theoretical Physics; School of Physics, University of the Witwatersrand, Wits 2050, South Africa. }
\author{Bruce Mellado}
\affiliation{University of the Witwatresrand, School of Physics,Private Bag 3, Wits 2050, South Africa. }
\author{Xifeng Ruan}
\affiliation{University of the Witwatresrand, School of Physics,Private Bag 3, Wits 2050, South Africa. }

%------------------------------------------------------------------------------
\begin{abstract} 
The signature produced by the Standard Model Higgs boson in the Vector Boson Fusion  (VBF) mechanism is usually pinpointed by requiring two well separated hadronic jets, one of which (at least) of them tends to be in the forward direction. With the increase of instantaneous luminosity at the LHC, the isolation of the Higgs boson produced with the VBF  mechanism is rendered more challenging. In this paper the feasibility of single jet tagging is explored in a high-luminosity scenario. It is demonstrated that the separation in rapidity between the tagging jet and the Higgs boson can be effectively used to isolate the VBF signal. This variable is robust from the experimental and QCD stand points. Single jet tagging allows us to probe the spin-CP quantum numbers of the Higgs boson.
\end{abstract} 
%------------------------------------------------------------------------------
\pacs{14.80.Cp, 13.60.-r, 11.30.Er} 
\maketitle

\section{Introduction}

The discovery of a Higgs boson~\cite{EWSB} by the ATLAS and CMS collaborations at the Large Hadron Collider (LHC)  is a major milestone for the history of particle physics~\cite{Aad:2012tfa,Chatrchyan:2012ufa}. It is also a new opportunity for deeper understanding of the fundamental interactions. A new sector is now available for exploration in the Standard Model (SM)~\cite{GSW} and physics beyond the SM (BSM). With the observation of the Higgs-like particle, the first measurements of the observables sensitive to its couplings to SM particles have become possible. The LHC experiments are expected to collect a sizable amount of Higgs boson candidates in the next few years. Together with searches for additional Higgs-like resonances, the exploration of couplings with increased statistics has become a focus. 

The exploration of couplings at the LHC suffers from a number of limitations. The measurement of the total width and life time is not possible. The width is too small to be measurable and a number of potential decay products would remain undetectable at the LHC. As a result, the LHC can only measure ratios of couplings in a quasi-model independent way~\cite{Zeppenfeld:2000td}. The isolation of the Higgs boson with the Vector Boson Fusion (VBF) mechanism is of great importance for the exploration of the coupling strength. The VBF mechanism is also critical to exploring the tensor structure of the $HVV$ couplings, where $V=W,Z$.

The isolation of the VBF mechanism with a large signal to background ratio and high purity is reliant on the ability to tag forward hadron jets. The ATLAS and CMS experiments have demonstrated the feasibility of forward jet tagging~\footnote{The acceptance of the ATLAS and CMS calorimeters lies in the range $\left|\eta\right|$, where $\eta=-\ln{\tan{{\theta \over 2}}}$.} in the challenging conditions of proton-proton collisions at the LHC. However,  with the increase of the instantaneous luminosity, necessary to reach O(100) fb$^{-1}$- O(1) ab$^{-1}$ integrated luminosity, the probability of fake jet tagging increases considerably. This leads experimentalists to increase the transverse momentum $\left(p_{T}\right)$ threshold, resulting in significant loss of signal acceptance. In a recent study for the assessment of the sensitivity to the VBF signal in the High-Luminosity (HL) LHC $p_T$  thresholds ranging from $50\gev$ to $77\gev$ have been considered, depending on the jet rapidity~\cite{ATL-PHYS-PUB-2014-012}. In this setup the expected accuracy of the VBF signal strength lags behind that of other measurements, such as the $WH$ and $ttH$ production mechanisms.  In order to ameliorate this problem it is suggested to revisit some of the ideas pertaining to isolating the Higgs boson with a single jet tag~\cite{pl_431_410,pl_611_60,pr_76_093007}. Single jet tagging was explored with the intention to identify regions of the phase-space where the Higgs boson could be isolated from non-resonant backgrounds. Here single jet tagging is re-evaluated with the primary intention to separate VBF from the gluon-gluon fusion (ggF) production mechanism.

In this paper the rapidity difference between the leading jet and the Higgs boson is considered as a means of achieving the necessary signal to background ratio with single jet tagging. Here it is demonstrated that this observable is robust from the QCD stand point for both VBF production and the ggF mechanisms. A perturbative analysis is performed to understand the stability of this observable against scale variations. Effects related to multiple soft gluon radiation are also investigated. The discriminating power of the observable studied here is evaluated in the context of the di-photon decay channel. The ability to study the spin-CP quantum numbers of the Higgs boson in the presence of a single jet tag is discussed. 

%In addition, the QCD uncertainties of the gluon-gluon fusion (ggF) production mechanism remain large enough so as to significantly affect the precision of future %measurements. The QCD scale uncertainties of the vector boson fusion, VBF, process are quite small. However, the contamination from radiative processes %ggF+$2j$ is sizeable and poses a limiting factor on the precision of the determination of the absolute rate of the Higgs boson production  via VBF. 

%The discussion on the measurement of coupling ratios has been recently renewed to add final states that had not been considered in the earlier %study~\cite{Djouadi:2012rh}.

% These mostly relate to the production of the Higgs boson in association with one high $p_T$ hadronic jet~\cite{pl_431_410,pl_611_60,pr_76_093007}.

The article is organized as follows: Section~\ref{sec:hplusjet} gives a brief overview of the Higgs boson production in association with high $p_T$ hadronic jets; Section~\ref{sec:ratprod} discusses ratios relevant to the $H+1j$ final state;  Section~\ref{sec:setup} gives a brief account of the tools used; Section~\ref{sec:QCD} reports the perturbative analysis of the observable under study; Section~\ref{sec:results} quantifies the discriminating power of the observable under study; Section~\ref{sec:spincp} discusses the ability to probe the spin-CP quantum numbers of the Higgs boson produced via VBF with single jet tagging; 
Section~\ref{sec:summary} summarizes the conclusions of the paper.  

\section{The Higgs boson and  jet production}
\label{sec:hplusjet}
 The phenomenology pertaining to the production of the Higgs boson at hadron colliders is vast and well understood~\cite{Djouadi:2005gi}. The leading production mechanism for Higgs bosons in association with high $p_T$ hadronic jets is the ggF mechanism, which occurs via a quark loop. In this process the production of jets involve radiative corrections. In the limit that the top quark is very heavy, the cross-section can be computed via an effective Lagrangian (see Ref.~\cite{Ravindran:2002dc} and references therein) as:
 \begin{equation}
 \mathcal{L}_{eff} =  - {1 \over 4 }A \Phi G^{A}_{\mu\nu}G^{A,\mu\nu},
 \end{equation}
where $\Phi$ stands for the scalar Higgs boson field and $G^{A}_{\mu\nu}$ is the field strength of the SU(3) color gluon field. The effective coupling $A=\alpha_s/(3\pi\nu)$, where $\nu=(G_F\sqrt{2})^{-1}=(246$\,GeV$)^2$ is the vacuum expectation value (VEV). The effective Lagrangian generates vertices leading to the production of the Higgs boson in association with gluons (the Feynman rules can be found in Ref.~\cite{ggF2jLO}, for instance). The leading process for the production of $H+1j$ emerges mainly from the partonic process
\begin{equation}
gg\rightarrow gH.
\end{equation}
The cross-section for $H+1j$ is known at $\alpha_s^4$~\cite{Schmidt:1997wr,deFlorian:1999zd,Ravindran:2002dc,Glosser:2002gm}. The production of $H+1j$  from $gg\rightarrow H+j$ is known at $\alpha_s^5$~\cite{Boughezal:2013uia}. Scale-driven variations of the cross-section are typically calculated by taking the largest variations by changing the renormalization ($\mu_R$) and factorization ($\mu_F$) scales by factors of two.  In this setup the cross-section varies within 20\% in a wide range of the $p_T$ of the leading parton relevant to Higgs boson searches at the LHC. The cross-section variation obtained by setting up the nominal scales to the Higgs boson mass, or to a dynamic choice of the Higgs boson transverse energy, are very similar. Using the effective Lagrangian approach, significant differences in the radiation patterns are observed with respect to Drell-Yan production~\cite{pl_431_410,pl_611_60}. 

The leading and subleading partonic processes for the production of $H+2j$ with ggF at the LHC are
\begin{equation}
gg\rightarrow ggH, ~qg\rightarrow qgH.
\end{equation}
The cross-secton for $H+2j$ with the ggF production mechanism is known at $\alpha_s^5$~\cite{Campbell:2006xx}. The lower order amplitudes for $H+1j$ and $H+2j$ scattering are available exactly, without the use of the effective coupling approach. These calculations are quite complex and one does not expect that higher order corrections in QCD will be calculated for the exact top mass. It needs to be argued that the two loop calculations performed for $H+1j$ and $H+2j$ scattering are valid for the Higgs mass, $m_{H}$, and transverse momentum, $p_{TH}$, smaller than the top mass. 

The Higgs boson production via the VBF is a sub-leading process that provides high $p_T$ hadronic jets at leading order (LO). The impact of the QCD higher order corrections on the production cross-section and the jet kinematics are known to be small. In order to appreciate the unique kinematics of the VBF process it is most intuitive to express the cross-section in a factorized form. Consider a fermion $f$ of a center-of-mass (c.m.)~energy $E$ radiating a gauge boson $V$ ($s \gg M_V^2$), the cross-section of the scattering $fa\rightarrow f^{\prime} X$ via $V$ exchange can be expressed as:
\begin{equation}
\sigma(fa\rightarrow f^{\prime} X) \approx \int dx\ dp_T^2\ P_{V/f}(x,p_T^2)\ \sigma(Va\rightarrow X),
\label{eq:effWapp}
\end{equation}
where $\sigma(Va\rightarrow X)$ is the cross-section of the $Va\rightarrow X$ scattering and $P_{V/f}$ 
can be viewed as the probability distribution for a weak boson $V$ of energy $xE$ and transverse momentum $p_T$.
The dominant kinematical feature is a nearly collinear radiation of $V$ off $f$, often called 
the ``Effective $W$-Approximation" (see Ref.~\cite{Han:2009pe} and references therein)
when the center of mass energy is much greater than the mass of the weak bosons, the
probability distributions of the weak bosons with different polarizations can be approximated by:
%~\cite{Cahn:1983ip,Chanowitz:1984ne,Kane:1984bb}:
\begin{eqnarray}
P_{V/f}^T(x,p_T^2) &\propto&
{1 + (1-x)^2 \over x } { p_T^2 \over \left(p_T^2 + (1-x)M_V^2\right)^2 }
\label{eq:PT} \\
P_{V/f}^L(x,p_T^2) &\propto&
{1-x \over x } { (1-x) M_V^2 \over \left(p_T^2 + (1-x)M_V^2\right)^2 }.
\label{eq:PL}
\end{eqnarray}
%where $P_{V/f}^T$ and $P_{V/f}^L$ are the probabilities for transversely and longitudinally
% polarized weak bosons,  respectively. 
%
These expressions lead to the following observations:
\begin{itemize}
\item[1]
Unlike the QCD partons that scale like $1/p_T^2$ at the low 
transverse momentum, the final state quark $f'$ typically has 
$p_T\sim\sqrt{1-x}M_V\leq M_W$.

\item[2]  Due to the $1/x$ behavior for the gauge boson distribution, the out-going 
parton energy $\left(1-x\right)E$ tends to be high. Consequently, it leads to an energetic  forward jet
with small, but finite, angle with respect to the beam.

\item[3] At high $p_T$, $P_{V/f}^T\sim 1/p_T^2$ and $P_{V/f}^L\sim 1/p_T^4$, and thus 
the contribution from the longitudinally polarized gauge bosons is relatively suppressed 
at high $p_T$ to that of the transversely polarized. 

\end{itemize}

In conclusion, the production of jets in association with the Higgs boson displays significant differences with respect to the production of jets in association with other particles in the SM. These differences are exploited when exploring the phase-space to isolate the Higgs boson signal. These features are also prominent in the production of the Higgs boson in association with one high transverse momentum jet.

\section{Production mechanism ratios and single jet tagging}
\label{sec:ratprod}

In this section the role of single jet tagging for the exploration of some of the properties of the Higgs boson is discussed. Given the limitations imposed by the inability to measure branching ratios in proton-proton collisions, it is convenient to define appropriate ratios. By defining ratios, where the rate of the Higgs boson decaying into the same flavor of particles is considered, uncertainties related to the total decay width cancel out. The ggF, VBF and VH production mechanisms are   sensitive to different couplings. In searching for physics beyond one can consider two groups of ratios:
\begin{itemize}
\item Ratios of rates of the same decay modes involving the production of ggF to VBF. If the VBF signal is isolated with the help of the $H+2j$ category, then we encounter a difficulty. QCD-related uncertainties of the contamination of the ggF process in the phase-space of the $H+2j$ category, used for the isolation of the VBF mechanism, would not cancel out. This would lead to approximately $15\%$ of theoretical uncertainty on the ratio. To estimate it one would need to add experimental uncertainties, which are significant here too. In this paper a ratio based on the $H+1j$ final state is suggested instead, as a means to secure strong cancelation of these effects.
\item Ratios of rates of the same decay modes involving the production of ggF to $VH$ are used. The isolation of $VH$ with a dedicated $H+2j$ category is hindered by the large contamination from the ggF mechanism. In order to effectively pursue a similar approach as suggested in Ref.~\cite{Butterworth:2008iy}, where the di-jet system is required to be boosted. This ratio would still suffer from similar theoretical uncertainties, as in the case discussed above. Another important ratio emerges from final states with leptons. Despite the reduced rate, the ratio involving leptons provides for an excellent opportunity to isolate the $VH$ production mechanism without concerns about contamination from the ggF mechanism. 
\end{itemize}
Based on this discussion let's consider the following experimental ratio:
\begin{equation}
R^{ggF}_{VBF}(1j)=  { g_2 + V_2  \over g_1 + V_1 } = { \xi^g(1j) g_1 + V_2  \over g_1 + V_1 } \approx   \xi^g(1j)  + {V_2  \over g_1},
\label{eq:ratprod1}
\end{equation}
where $g_1$ ($V_1$) and $g_2$ ($V_2$) correspond to the rate of the Higgs boson via the ggF (VBF) mechanism in the region of the phase-space enriched with the ggF (VBF) mechanism. Here $g_1$ is the experimental measurement of the rate of the ggF$+1j$, whereas $g_2$ and $V_2$ would be estimates extracted with the procedure.The region of the phase-space where the ggF mechanism dominates over the VBF is where QCD-like radiation patterns are characteristic. Theory uncertainties from $\xi^g(1j)=g_2/g_1$ and $V_2$ would need to be considered. The dominant theory uncertainty would emerge from the QCD uncertainties of the rate of ggF$+1j$, whereas experimental uncertainties would tend to cancel out. It is important to note that the ratio $R^{ggF}_{VBF}(1j)$ is robust against pile-up effects. The ratio $V_1/g_1$ is expected to be small, hence the approximation in Eq.(\ref{eq:ratprod1}).  The following expression is used for the relative uncertainty of the extraction of the VBF signal in the $H+1j$ category:
\begin{equation}
\left(\sqrt{g_2+V_2} \oplus \sqrt{g1}\xi^g(1j) \oplus \delta\xi^g(1j) g_1 \right) / V_2,
\label{eq:ratprod1err}
\end{equation}
where $\delta\xi^g(1j)$ is the scale variation obtained with the next-to-leading-order (NLO) matrix element of the ggF+$1j$ process. The first two terms in Eq.(\ref{eq:ratprod1err}) are related to the statistical error of the measurement. It is found that the best discriminator to disentangle the ggF and VBF processes is the rapidity difference between the Higgs boson and the leading jet, $\Delta y_{Hj}$; the separation in rapidity between the Higgs boson and the leading jet. For a quantitative statement see Sec.~\ref{sec:results}.
%Figure~\ref{fig:2} shows the  $\Delta y_{Hj}$ for ggF+$1j$  and VBF after requiring $p_{TH}>100\,$GeV. 

%Results of the relative uncertainty are shown in Fig.~\ref{fig:3} for 300\,fb$^{-1}$ of integrated luminosity and assuming SM rates of the Higgs %boson production. Results are shown for three cuts on $p_{TH}$. The dominant contribution comes from the statistical uncertainties, where the %contribution from the third term is small. The value of $\delta\xi^g(1j)$ for an optimal value of $\Delta y_{Hj}>1.5$ is $\approx 5\%$. 

%\begin{figure}[t]
 %\begin{center}
 %\includegraphics[width=9cm]{./Figures/detaj1hcut2VG.eps}
 %\end{center}
 %\caption{Expected distribution of $\Delta y_{Hj}$ for ggF+$1j$ (black) and VBF (red) after requiring $p_{TH}>100\,$GeV. Results are shown after %requiring two photon with $p_T>30\,$GeV in the range $\left|\eta\right|<2.4$. \label{fig:2}
%}
%\end{figure}

%\begin{figure}[t]
% \begin{center}
% \includegraphics[width=9cm]{./Figures/optdeta.eps}
% \end{center}
% \caption{Relative error on the extraction of $V_2$ (see Expression~\ref{eq:ratprod1err} with the $H+1j$ category. Results are obtained for $p_{TH}>50, 75, 100\,$GeV corresponding to the solid, dashed and dotted curves, respectively. Results are obtained for 300\,fb$^{-1}$ of integrated luminosity and assuming SM rates of the Higgs boson production. \label{fig:3}
%}
%\end{figure}

This approach is valid for the extraction of the VBF signal that later can be related to the quasi-inclusive rate of ggF, $g_0$:
\begin{equation}
R^{ggF}_{VBF}=  { g_2 + V_2  \over g_0 + V_0 } = { \xi^g(1j) g_1 + V_2  \over g_0 + V_0 } \approx   \xi^g(1j) {g_1 \over  g_0} +  {V_2  \over g_0},
\label{eq:ratprod2}
\end{equation}
where $g_0=g_{incl}-g_1$ and $g_{incl}$ would be the total inclusive cross-section for the ggF mechanism. The dominant theoretical error in this case would be the QCD uncertainty in the total cross-section of the ggF mechanism. In this approach the theory uncertainty on the ggF+$1j$  rate would cancel out, except for its contribution to $g_0$, which is small. Experimental uncertainties related to hadronic jet reconstruction would not cancel out in this approach. That said, since the VBF signal is extracted with a $H+1j$ category, these uncertainties are not expected to be large. 

Third approach would be to use the $H+2j$ category for the extraction of the VBF signal. Here $g_2 =  \xi^g(2j) g_2^{\prime}$, where $g_2^{\prime}$ would be measured in the QCD-like region. This can be achieved by applying requirements on the rapidity difference between the tagging jets ($\Delta y_{jj}$) in order to define a region with dominant  ggF contribution and another with dominant VBF contribution. 

\section{Setup and tools}
\label{sec:setup}

%%%%%%%%%%%%
%% POWHEG %% 
%%%%%%%%%%%%
Monte-Carlo events were generated for the two production modes: ggF and VBF.  
%The ggF events were generated using the {\tt POWHEG}  ggF$+2j$ package~\cite{Campbell:2012am}, and are accurate to NLO for Higgs produced in association with up to two jets. %HJJ was used in association with {\tt MINLO} (Multi-scale improved NLO)~\cite{MINLO,Hamilton:2012np}, 
%which uses the CKKW prescription~\cite{Catani:2001cc,Krauss:2002up,Mrenna:2003if}
%which sets the emission scales to the $p_T$ of the emitted partons. 
Two versions of the {\tt MINLO}~\cite{MINLO,Hamilton:2012np} generator were used for the production of ggF+jets: HJ and HJJ. The first incorporates NLO matrix elements up to one parton, whereas the second is up to two partons. 
The VBF events were also produced with {\tt POWHEG}~\cite{powheg_VBF} and are also accurate at NLO.
Both samples were produced for a Higgs mass of 126.8 GeV using the CT10 parton distribution functions at NLO~\cite{PDF:CT10}.

%%%%%%%%%%%%
%% PYTHIA %%
%%%%%%%%%%%%
The samples were then interfaced with {\tt Pythia8}~\cite{pythia8}	
which adds the showering and hadronization of the events as well as the underlying event.
Within {\tt Pythia8}, stable particles are clustered into jets using the anti-$k_\mathrm{t}$ algorithm~\cite{JetAlgo} with a cone size of $\Delta R=0.4$.

%%%%%%%%%%%%%%%%%%%
%% FIDUCIAL CUTS %%
%%%%%%%%%%%%%%%%%%%
A number of fiducial cuts were applied to the samples.  The $p_T$ of the leading (subleading) photon is required to be greater than 40 (30) GeV and within a rapidity $\left|y\right|<2.4$.  In addition, the photons are isolated, which is accomplished by requiring the amount of transverse energy within $\Delta R=0.4$ to be less than 14 GeV.  Jets are then required to have $p_T>30\,$GeV and to be within a rapidity $\left|y\right|<4.4$. An overlap removal is applied on jets, where any jet within $\Delta R<0.4$ of a photon is removed, and any jet within $\Delta R<0.2$ of an electron is removed.
% electrons are also isolated and have pT>15 GeV and |y|<2.47

The package {\tt MCFM}~\cite{Campbell:2006xx,Campbell:2010cz} is used for the evaluation of the scale uncertainties of ggF+jets and VBF in the corners of the phase-space of interest here (see Section~\ref{sec:QCD}). It is worth noting that while {\tt MCFM} is a parton-level generator, studies performed with {\tt POWHEG}   and {\tt MINLO}  are at particle level.

\section{Perturbative analysis}
\label{sec:QCD}

The scale variations of ggF+$1j$ and VBF are evaluated with {\tt MCFM} at NLO as a function of the rapidity difference between a Higgs boson and the leading jet at parton level. The K-factors~\footnote{The K-factors are defined as the ratio of the prediction at Next-to-Leading order to that of Leading order. These are computed with the same scales.} for ggF+$1j$  are remarkably flat up to $\Delta y_{Hj}\approx 5$, beyond which statistical fluctuations become a limiting factor. The scale variations are evaluated for the renormalization and factorization scales both at the same time, and separately. The size of the scale variations is also stable  for $\Delta y_{Hj}<5$. Scale variations for the VBF process are well behaved. However, unlike the ggF+$1j$ case, the K-factors are not flat with $\Delta y_{Hj}$. The K-factors behave almost linearly with $\Delta y_{Hj}$ ranging from 0.9 at $\Delta y_{Hj}\approx 0$ to 1.35 at $\Delta y_{Hj}=7$.	

It is important to note that the observable $\Delta y_{Hj}$ displays similar features to the invariant mass of the Higgs boson and the leading jet in terms of the flatness scale uncertainties. That said, experimentally, $\Delta y_{Hj}$ is  robust  with respect to hadronic energy scale uncertainties.

\begin{figure}[t]
 \begin{center}
 \includegraphics[width=9cm]{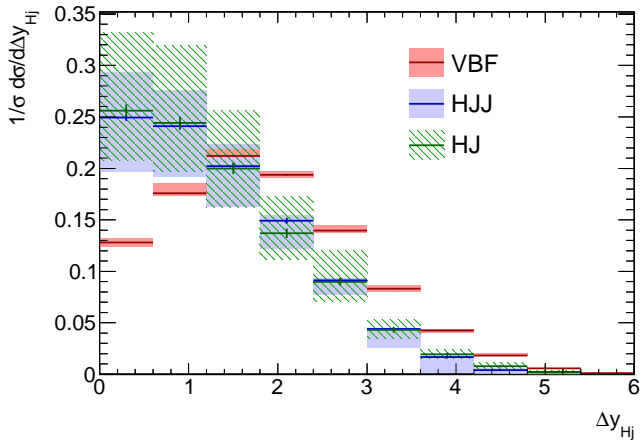}
 \end{center}
 \caption{Distributions of the rapidity separation between the Higgs boson and the tagging jet. The VBF and ggF$+1j$ processes are described with {\tt POWHEG} and {\tt MINLO}, respectively. The ggF$+1j$ production mechanism is described with the HJ and HJJ versions of {\tt MINLO} (see text). Variations in the differential cross-sections due to scale variations are shown in the form of bands around the central values.  \label{fig:detahj}
}
\end{figure}

Figure~\ref{fig:detahj} displays the $\Delta y_{Hj}$ distribution for the VBF and ggF$+1j$ production mechanisms at particle level (see Section~\ref{sec:setup}) .  Changes in the differential cross-sections due to scale variations are shown in the form of bands around the central values.  The scale variations for VBF are well behaved, as expected. 

The situation with ggF+jets requires some discussion. At low values of $\Delta y_{Hj}$ the cross-section variations due to scale variations are larger for HJ than for HJJ, which is expected. However for $\Delta y_{Hj}>3.5$ the cross-section variations are larger for HJ than for HJJ. This seems an indication that the calculation may not be particularly reliable for large values of $\Delta y_{Hj}$. Fortunately,  this region of the phase-space does not play a critical role in the separation between ggF and VBF. In Section~\ref{sec:results} it will be seen that a cut  of $\Delta y_{Hj}>1.4$ is an optimal requirement to separate ggF and VBF.  This requirement is far enough from what seems to be a problematic region. To obtain an estimate of the scale uncertainty for ggF events which fall into the $H+1j$ category, the ratio of events with $\Delta y_{Hj}>1.4$ to events with $\Delta y_{Hj}<1.4$ is studied.  The scale uncertainties for the ggF cross-section were found by varying the factorization and renormalization scales up and down by a factor of 2.  This uncertainty on the cross-section was then propagated to the $\Delta y_{Hj}$ ratio of events, and was found to be 6.5\%.

\section{Discrimination}
\label{sec:results}

In this Section a qualitative statement is made about the relevance of $\Delta y_{Hj}$ as a discriminator to extract the VBF signal. A generic corner of the phase-space is identified in order to evaluate loss of VBF signal acceptance as a result of the increase of the jet $p_T$ thresholds imposed by the pile-up conditions. The region defined by the following requirements assumes the presence of two hadronic jets with $p_T>30\,\gev$, and in the pseudorapidity range $\left|\eta_{j} \right|<4.5$: pseudorapidity separation between the tagging jets $\Delta\eta_{jj} > 2.8$, $\eta_{\gamma\gamma}  - \frac{\eta_{j1}+\eta_{j2}}{2} < 2.4$, where the indices indicate the object for which pseudorapidity is calculated, the azimuthal angle difference between the system of the tagging jets and the di-photon system, $ \Delta\phi_{\gamma\gamma, jj} > 2.6\,$rad, and the di-jet invariant mass, $m_{jj}>560\,\gev$. This region is best suited for the extraction of the VBF signal in the presence of two high $p_T$ jets. Two classes of Higgs boson events are identified: { \it double tag}, or events that pass the requirements specified above; { \it single tag}, or events that fall outside the region but that display a jet in the event with $\Delta y_{Hj}$ above a certain threshold. Events classified as {\it single tag}  appear in a region of the phase-space currently not explored for the extraction of the VBF signal by the ATLAS and CMS experiments.

\begin{figure}[t]
 \begin{center}
\includegraphics[width=9.cm]{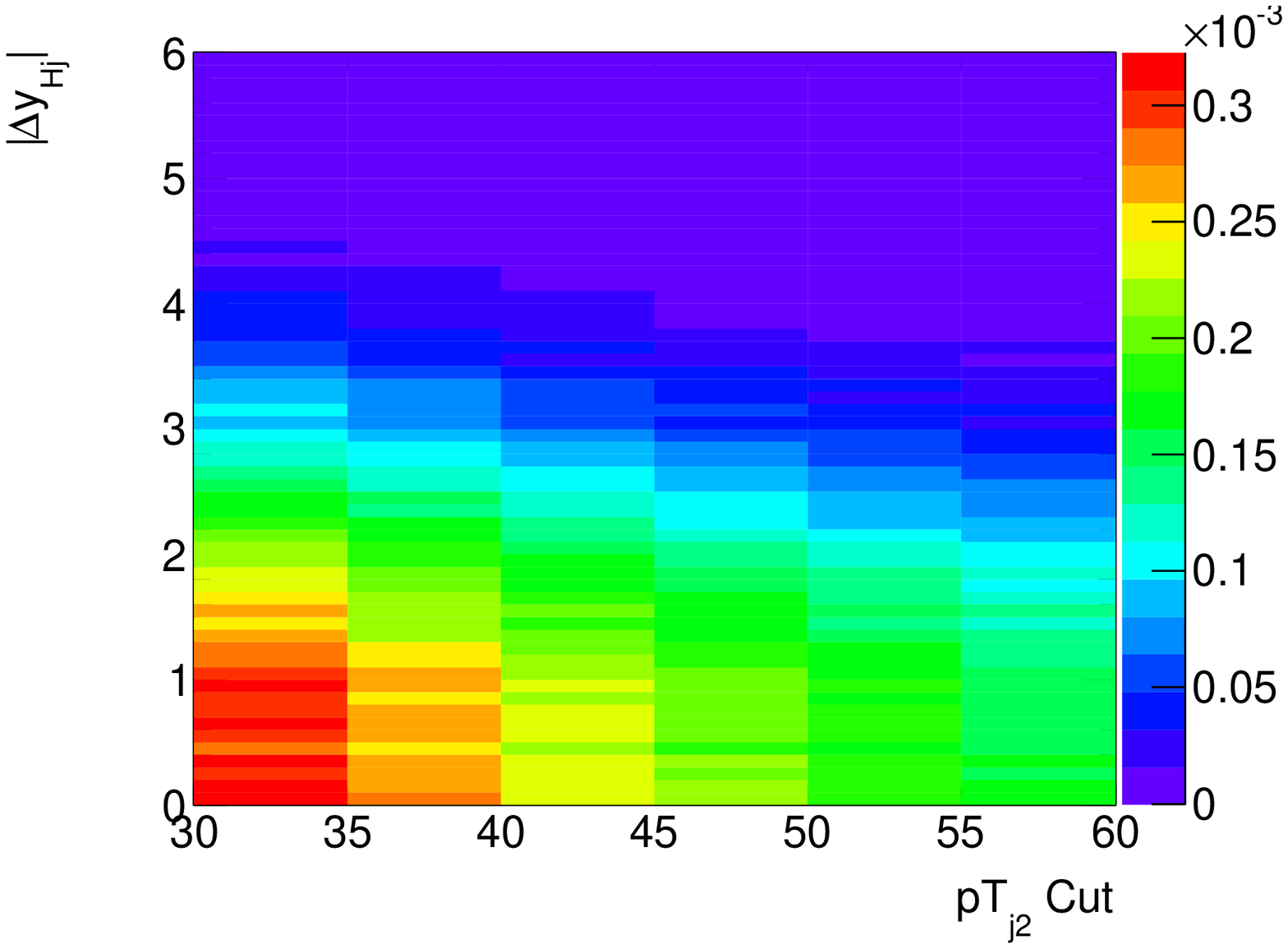}
\includegraphics[width=9.cm]{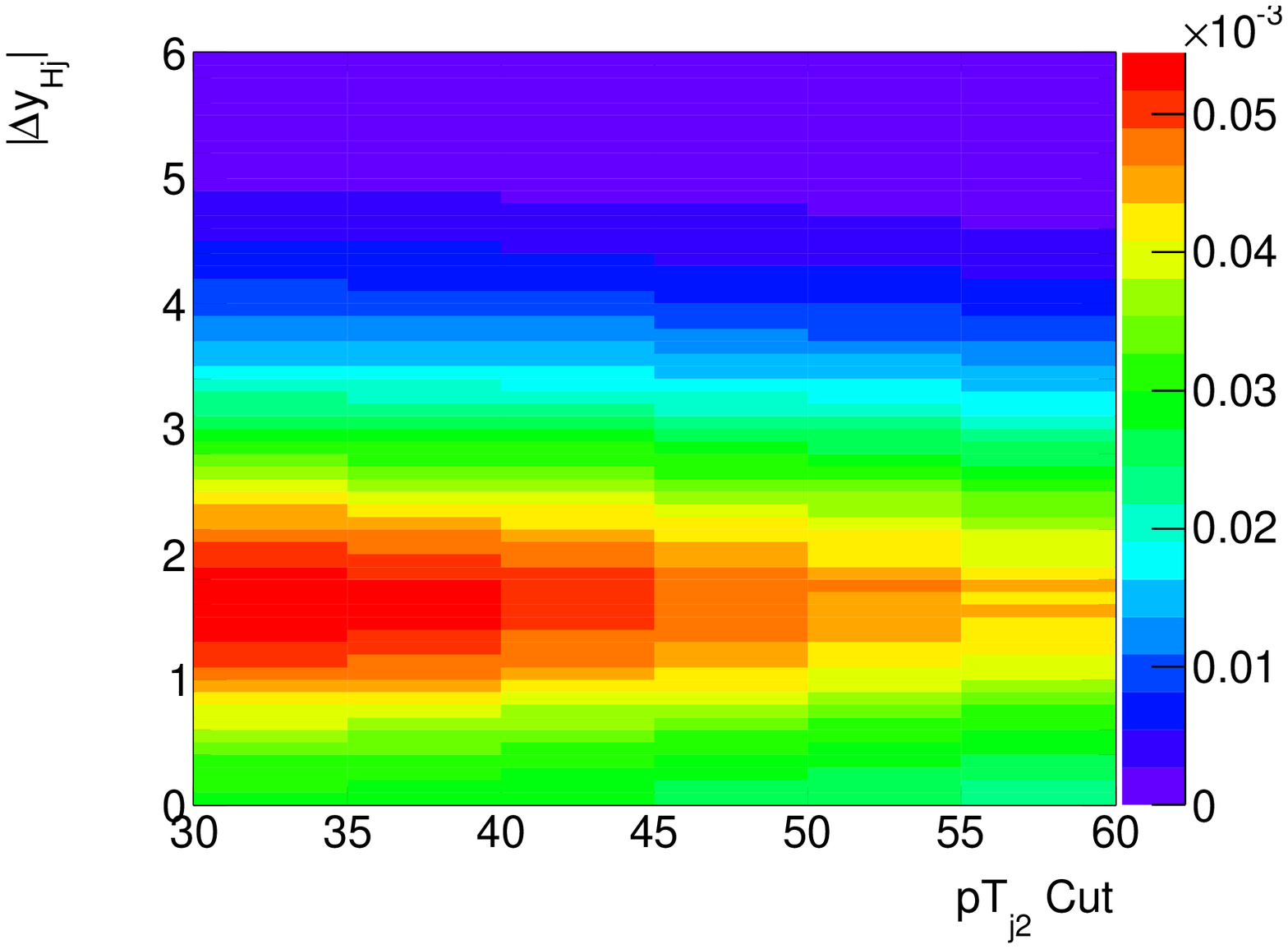}
 \end{center}
 \caption{Effective cross-section  (in pb) of ggF+jets and VBF production as a function of the jet $p_T$ and $\Delta y_{Hj}$ thresholds for {\it double tag} events with the di-photon decay channel. The upper and lower plots correspond to  ggF+jets and VBF production, respectively.  \label{fig:2D}
}
\end{figure}

It is important to evaluate the evolution of the Higgs boson signal cross-section with the jet $p_T$ threshold and its correlation with the  $\Delta y_{Hj}$. 
Figure~\ref{fig:2D} shows the dependence of the effective cross-section for the ggF+jets (upper plot) VBF (lower plot)  and  processes as a function of the sub-leading jet $p_T$ threshold, $p_{Tj2}$ and $\Delta y_{Hj}$ for double tag events. The correlation between $\Delta y_{Hj}$ and $p_{Tj2}$ is significantly different for both processes: whereas $\Delta y_{Hj}$ decreases with the $p_{Tj2}$ threshold for ggF+jets, one observes a weak correlation in VBF. This is an important feature for the effectiveness of  $\Delta y_{Hj}$ as a discriminator to extract the VBF signal: as the jet threshold increases the separation becomes stronger. 

Table~\ref{tab:results} displays the results of a one-dimensional optimization using $\Delta y_{Hj}$ as a discriminator, for different values of the jet $p_{T}$ threshold. The value of the threshold on $\Delta y_{Hj}$ depends little on the $p_T$ threshold and it is fixed at $\Delta y_{Hj}>1.4$.
Shown are the signal VBF and ggF+$1j$ background rates and the expected accuracy on the measurement of the VBF signal strength. For the evaluation of the latter a 5\% uncertainty on the ggF+$1j$ background extraction is assumed (see Section~\ref{sec:ratprod} for a discussion on the subject). The optimal value of the threshold of $\Delta y_{Hj}$ depends weakly on the jet $p_T$ threshold. For thresholds on $p_{Tj2}$ above $50\,\gev$ the sensitivity of the single tag category becomes dominant. 

The signal rate for both VBF and ggF+jets mechanisms for {\it double tag} events decrease rapidly with the jet $p_T$ threshold. This effect is further quantified in Tab.~\ref{tab:results}, where event yields for the VBF signal and ggF+jets rates for 300 fb$^{-1}$ integrated luminosity are given as a function of the jet $p_T$ threshold. When shifting the threshold from 30\,\gev \, to 55\,\gev,  the rate of VBF signals classified as {\it double tag} events drops by a factor of two. This effect seriously affects the sensitivity of the experiments to the extraction of the VBF in high instantaneous luminosity scenarios. The mild increase in the VBF signal to ggF+jets rate does not compensate the strong  loss of VBF signal. It is important to note that the effects discussed here are only applicable to the SM (see Section~\ref{sec:spincp}).

Table~\ref{tab:results} also displays the expected yield for the VBF production mechanism using {\it single tag}  events. The rate of this class of events evolves  with the increase of the $p_T$ threshold as a result of two competing effects: increase of yield that do not pass the {\it double tag} requirements and the decrease of yield because of increase of threshold on the $p_T$  of the leading jet. This leads to a significantly milder decrease in the yield with the $p_T$ threshold for {\it single tag} events compared to that of {\it double tag} events. The rate of VBF to ggF yields is significantly poorer for {\it single tag} events compared to {\it double tag} events. This is partially alleviated by the large VBF signal yield produced by the {\it single tag} category.

\begin{table}[t]\scriptsize
  \begin{tabular}{c||cccccccc||c}
  \hline\hline
& & \multicolumn{3}{c}{Double tag} & & \multicolumn{3}{c}{Single tag} \\

 $p_{Tj}$ &	 $S$ & ggF & $S$/ggF & $\Delta\mu$ & $S$ & ggF & $S$/ggF & $\Delta\mu$ & $\Delta\mu^{Tot}$ \\
  \hline\hline 
30 &  137 & 30.7	 & 4.48 & 0.095 & 262 & 1204 & 0.217 & 0.146 & 0.080 \\
35 &  120 & 25.3 & 4.75 & 0.101 & 256 & 989 & 0.259 & 0.138 & 0.081 \\
40 &  103 & 21.0 & 4.92 & 0.108 & 247 & 815 & 0.304 & 0.132 & 0.084 \\
45 &  87.6& 17.3 & 5.07 & 0.117 & 236 & 681 & 0.346 & 0.128 & 0.087 \\
50 &  73.6 & 14.8 & 4.96 & 0.128 & 222 & 574 & 0.387 & 0.127 & 0.090 \\
55 &  61.2 & 12.1 & 5.04 & 0.140 & 208 & 483 & 0.431 & 0.126 & 0.094 \\
%30 & 1.4 & 137.321 & 30.668 & 4.478 & 0.095 & 261.526 & 1203.739 & 0.217 & 0.146 & 0.080 \\
%35 & 1.4 & 120.202 & 25.320 & 4.747 & 0.101 & 255.874 & 988.994 & 0.259 & 0.138 & 0.081 \\
%40 & 1.4 & 103.333 & 20.985 & 4.924 & 0.108 & 247.336 & 814.661 & 0.304 & 0.132 & 0.084 \\
%45 & 1.4 & 87.569 & 17.285 & 5.066 & 0.117 & 235.912 & 681.020 & 0.346 & 0.128 & 0.087 \\
%50 & 1.4 & 73.577 & 14.827 & 4.962 & 0.128 & 222.203 & 573.959 & 0.387 & 0.127 & 0.090 \\
%55 & 1.4 & 61.162 & 12.139 & 5.039 & 0.140 & 207.925 & 482.780 & 0.431 & 0.126 & 0.094 \\
%30 & 1.4 & 137.3 & 30.7 & 4.48 & 0.095 & 261.5 & 1203 & 0.22 & 0.146 & 0.080 \\
%35 & 1.4 & 120.2 & 25.3 & 4.75 & 0.101 & 255.9 & 989.0 & 0.26 & 0.138 & 0.081 \\
%40 & 1.4 & 103.3 & 30.0 & 4.92 & 0.108 & 247.3 & 814.7 & 0.30 & 0.132 & 0.084 \\
%45 & 1.4 & 87.6 & 17.3 & 5.07 & 0.117 & 235.9 & 681.0 & 0.35 & 0.128 & 0.087 \\
%50 & 1.4 & 73.6 & 14.8 & 4.96 & 0.128 & 222.2 & 574.0 & 0.39 & 0.127 & 0.090 \\
%55 & 1.4 & 61.2 & 12.1 & 5.04 & 0.140 & 207.9 & 482.8 & 0.43 & 0.126 & 0.094 \\
  \hline\hline
  \end{tabular}
\caption{VBF signal and ggF+jets rates for 300 fb$^{-1}$ integrated luminosity with the di-photon decay channel Thresholds on the jet $p_T$ are given in $\gev$. Results for different values of the threshold on $p_{Tj2}$ are given and are obtained for an optimal requirement of $\Delta y_{Hj}>1.4$. Results are shown for the expected accuracy of the signal strength measurement for the individual categories and their combination, $\Delta\mu^{Tot}$. \label{tab:results}}
\end{table}

%\begin{figure}[t]
% \begin{center}
%% \includegraphics[width=9cm]{./Figures/VBF_nonVbfTopo_1Jcateogry.eps}
% \includegraphics[width=9cm]{./Figures/VBF_vbfTopo_in1Jcategory.eps}	
% \end{center}
% \caption{Effective cross-sections (in pb) as a function of the jet $p_T$ threshold and $\Delta y_{Hj}$ for VBF signal  production using Single tag events (see text).  \label{fig:2D_VBF}
%}
%\end{figure}

A study of the $\Delta y_{Hj}$ spectrum displayed by the di-photon non-resonant production was studied with the {\tt SHERPA} package~\cite{Gleisberg:2008ta}. The shape of the $\Delta y_{Hj}$ spectrum follows closely that of ggF+jets.

\section{Exploration of Spin-CP quantum numbers}
\label{sec:spincp}

In Ref.~\cite{Plehn:2001nj} it was suggested to explore the spin-CP quantum numbers of the Higgs boson in VBF via the study of the azimuthal angle correlation of the scattered quarks. Experimentally this implies reconstructing two well separated hadronic jets. It is difficult to gain indirect access to this observable in the final state considered here. The azimuthal angle separation between the Higgs boson and the leading jet does not have sufficient sensitivity to the information of interest. In Refs.~\cite{Englert:2012xt,Djouadi:2013yb} it was pointed out that the tensor structure of the $HVV$ vertex ($V=Z,W$) manifests itself through other observables in addition to the one considered in Ref.~\cite{Plehn:2001nj}. The sensitivity to new physics in the $HVV$ couplings in the $\Delta y_{Hj}$ distribution is evaluated here.

In the SM, the couplings of the Higgs boson to the massive electroweak gauge 
bosons are precisely formulated and come out as  $g_{HVV}
\propto  g M_V V_\mu V^\mu$ where $g$ is the SU(2) coupling constant.  However, this
is not the most general form of the Higgs--gauge boson vertex. Parametrising the
coupling of a scalar state to  two vector bosons  in the form $i\Gamma^{\mu\nu}
(p,q) \epsilon_\mu(p) \epsilon^\ast_\nu(q)$, one can write down the most general
form of the $HVV$  vertex as $\Gamma_{\mu\nu}(p,q)= \Gamma^{\rm  SM}_{\mu\nu}
+\Gamma^{\rm BSM}_{\mu\nu}(p,q)$, with the SM and the beyond SM  components
given by:

\begin{equation}
\Gamma^{\rm  SM}_{\mu\nu} = - gM_V\, g_{\mu\nu} \label{eqn:SMvertex}, 
\end{equation}
\begin{equation}
\Gamma^{\rm BSM}_{\mu\nu}(p,q) = \frac{g}{M_V}\left[ 
\lambda \left(p \cdot q\, g_{\mu\nu} - p_\nu q_\mu  \right)
+ \, \lambda^\prime\ \epsilon_{\mu\nu\rho\sigma}p^\rho q^\sigma \right],
\label{eqn:BSMvertex}
\end{equation}
where $\lambda$ and $\lambda^\prime$ are effective  coupling strengths,
respectively for higher dimension CP-even and CP-odd operators, and
we will assume that they are the same for $W$ and $Z$ bosons. These operators
may be generated within the SM at higher orders of perturbation theory, 
although the resulting couplings are likely to be very small. In
general, $\lambda$ and $\lambda^\prime$ can be treated as momentum dependent
form factors that may also be complex valued. However, we take the approach
that  BSM vertices can be generated from an effective
Lagrangian, which treats $\lambda$ and $\lambda^\prime$ as coupling
constants~\cite{Plehn:2001nj}.  The most striking difference between the SM and BSM
vertices of Eqs.~(\ref{eqn:SMvertex}) and (\ref{eqn:BSMvertex}) is that the 
latter has an explicit dependence on the momentum of the gauge bosons. It is
this feature that is the source of the differences that the  BSM vertices
generate in the kinematic distributions of tagging jets in the VBF and $VH$
processes, compared to the SM case.

In our analysis, the vertices for the Lagrangians in the SM and in BSM with
spin-0 bosons are calculated  in {\scshape{FeynRules}}~\cite{FEYNRULES} and
passed to the event-generator {\scshape{MadGraph}}~\cite{Alwall:2011uj}, which is used for the
generation of  the matrix elements for Higgs production in VBF. To
obtain the cross-sections and distributions at parton-level, the CTEQ6L1
parton distribution functions  are used~\cite{CTEQ}. The factorization and re-normalization scales are set on an event-by-event
basis to the transverse energy of the Higgs boson. 
For the selection cuts, 
partons are required to have
transverse momentum $p_T>10\,$GeV, rapidity $\left|y\right|<5$ and be separated
by $\Delta R>0.4$. %The parton distribution function CTEQ6.6 is used.

\begin{figure}[t]
 \begin{center}
\includegraphics[width=9cm]{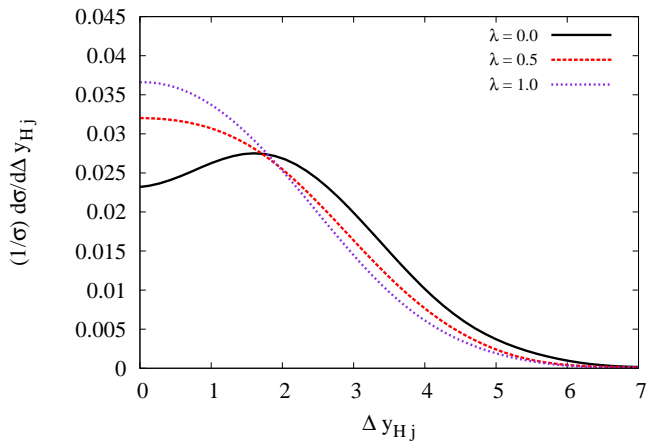}
 \end{center}
 \caption{Distribution of the rapidity separation between the Higgs boson and the leading jet in VBF.  Results are shown at parton level for the SM case ($\lambda=0$) and non-zero BSM contributions (see text). \label{fig:dyhj1_BSM}
}
\end{figure}

Figure~\ref{fig:dyhj1_BSM} displays the rapidity separation between the Higgs boson and the leading parton in the event. Results are shown at parton level. The solid black curve corresponds to the SM case, when $\lambda = \lambda^{\prime} = 0$. The dotted and dashed line include admixtures of the SM and BSM contributions with $\lambda^{\prime} = 0$ and $\lambda = 1, 0.5$, respectively.
As pointed out in Refs.~\cite{Englert:2012xt,Djouadi:2013yb}, the BSM vertexes in Eq.(\ref{eqn:BSMvertex}) introduce dependence on the particle momenta. This feature distorts the kinematics of the scattered quarks with respect to the prediction of the SM. One of the relevant effects is the reduction of the rapidity separation between the scattered quarks. Figure~\ref{fig:dyhj1_BSM} illustrates the effect on  the rapidity separation between the Higgs boson and the leading jet. With the inclusion of spin-0$^{+}$ BSM admixtures, the $\Delta y_{Hj}$ distribution is pushed towards lower values. The jet transverse momentum distribution is also a potential discriminant to explore the tensor structure of the $HVV$ coupling.

\section{Conclusions}
\label{sec:summary}

% BM editing
With the increase of the number of soft proton-proton collisions at the LHC, the probablity for fake forward jets will increase significantly. As a result jet transverse momentum thresholds will need to be increased, strongly reducing the phase-space to isolate the Higgs boson produced with the VBF mechanism using two well separated hadronic jets. The prospects of isolating the VBF mechanism with single jet tagging is explored here by using the difference in rapidity between the leading jet and the Higgs boson as a discriminator. It is demonstrated that this observable is robust from the QCD standpoint for both the VBF and ggF production mechanisms.   For thresholds of the jet transverse momenta greater than 50\,\gev, the sensitivity to the VBF mechanism of the {\it single tag} final state may become dominant. The combination of the {\it single} and {\it double} tagged final states provides enhanced stability of the measurement of the Higgs boson rate produced via VBF against stringent pileup conditions. The exploration of the Higgs boson spin-CP quantum numbers via VBF is not only possible with double jet tagging. Here it is demonstrated that the spin-CP quantum numbers can also be explored with the VBF mechanism using single jet tagging.

\end{document}